# Quantifying Gait Changes Using Microsoft Kinect and Sample Entropy

**Behnam Malmir, Shing I Chang, Malgorzata Rys & Dylan Darter**

*Department of Industrial and Manufacturing Systems Engineering, Kansas State University*
*Manhattan, KS 66506, USA*

## Abstract

This study describes a method to quantify potential gait changes of human subjects. Microsoft Kinect devices were used to provide and track coordinates of fifteen different joints of a subject over time. Three male subjects walk a 10-foot path multiple times with and without motion-restricting devices. Their walking patterns were recorded via two Kinect devices through frontal and sagittal planes. A modified sample entropy (SE) value was computed to quantify the variability of the time series for each joint. The SE values with and without motion-restricting devices were used to compare the changes in each joint. The preliminary results of the experiments show that the proposed quantification method can detect differences in walking patterns with and without motion-restricting devices. The proposed method has a potential to be applied to track personal progress in physical therapy sessions.

## 1. Introduction

Analyzing people's walking patterns provides essential information for measuring progress of physical therapy. Microsoft Kinect device may provide a mean to capture walking pattern of a person. Many researchers have used 3D kinematic data obtained by Microsoft Kinect devices in measuring activities and postural analysis studies. Eltoukhy *et al.* [1] used Kinect to identify different patients. Their study showed gait patterns differed between healthy people and those with Parkinson's disease. Xu *et al.* [2] tracked shoulder movements while a person was using a computer in an effort to reduce injuries. Their study found placing Kinect to the front of participants yielded more accurate shoulder measurements than placing the camera 15 to 30 degrees to the side. Vernon *et al.* [3] examined the test-retest reliability measures of some other kinematic measures, such as step length and stride length, to determine whether they could improve prediction performance in common clinical tests.

Quantification is essential for tracking joint health, especially for individuals who are undergoing physical therapy and are affected by an age-related disability. This quantification strategy can be extended to other physical health related areas. For example, it can help monitor elevated risk of falling as reflected in gait changes due to physical weakness. Sosnoff *et al.* [4] examined gait changes in persons with multiple sclerosis (MS) who had minimal disability, in doing so they found out that persons with MS walked with fewer, shorter, and took wider steps than healthy individuals. Those characteristics provided much more identifiable differences in walking patterns than age and gender. These patients also had a greater variability in the time between steps.

This study aims to quantify human gait changes using human skeleton coordinates recorded over time. The proposed procedure can be applied to fall prediction of elderly people, physical therapy, and sport science. Hondori and Khademi [5] studied the technical and clinical impacts of Kinect on physical therapy and rehabilitation. Subjects included in this research include elderly patients with neurological disorders, strokes, Parkinson, cerebral palsy, and MS.

### 1.1 Problem Statement

This study explores the possibility of quantifying gait changes of a human subject. The proposed method summarizes walking patterns over time using human skeleton coordinates derived from two Kinect devices. Gait is a key factor in determining the overall health of a subject. Therefore, the creation of a personal gait profile would be helpful in tracking personal well-being, particularly for the elderly population. For example, large changes in an elderly person's gait profile may be an indication of elevated fall risk. These profiles can also be used to track progress in a series of physical therapy sessions.

### 1.2 Experimental Equipment

Kinect contains an RGB camera, a depth sensor, and a multiarray microphone. Kinect's depth sensor can capture 3D data and does not require any particular lighting for the system, allowing it to capture data indoors or outdoors. In this study, however, we used the same room with the same lighting throughout the experimental period. Kinect is capable of a frame rate of 9–30 fps and a resolution of 640 x 480 that can be increased to 1280 x 1024 using a lower frame rate [5]. In this study, Kinect was set to record 30 fps of all joints in both directions, as shown in Figure 1. A customized software based on software development kit (SDK) was developed in C# language to gather data under a Windows operation system. A dynamic link library (DLL) was used to obtain the coordinates of selected skeleton joints in three different axes (X, Y, Z) = (anteroposterior, vertical, and mediolateral).

### 1.3 Hypothesis

The first question behind this study is whether Kinect can create a personalized walking profile by recording and tracking skeleton joint position data. This entails accuracy, precision, reliability, and ease of use of the system. All these factors are being considered to determine if this system is a practical choice for gait analysis. Through literature survey and our own experiments [6], we can confirm that Kinect device is capable of generating consistent results in terms of the proposed sample entropy method. Another question is whether this profile based on individual joints can be used in quantifying changes in human gait instead of gait parameters that are widely used in many studies [7-10]. The answer to this question is the main subject of this study.

## 2. Experimental Setup, Plan, and Data Collection

In this study, a simple walk test was conducted on three healthy subjects. The layout of the testing area is shown in Figure 1, in which two camera angles are used to record subjects' walking data. The testing area includes a 10-foot path that each subject walks through, turns around, and walks back. One Kinect camera is placed eight feet away from the walking path to the side and the other one is placed in front and three feet away from the end point.

All three subjects were instructed to wear athletic shoes, fitting shorts, and a t-shirt to ensure that the system can capture the joints accurately. The subjects were also asked to repeatedly wear similar clothing that would not interfere with the results of the experiment. The test subjects were instructed when to begin the test and a physical marker was placed near the end of the test area, so the subjects were aware of where they needed to stop without having to look down, which we found to skew results. The test subjects were also asked to lead with the same foot for each walk. They were instructed to walk at a consistent pace. Operators of Kinects counted down from three to ensure cohesive recording of test subjects.

Through the proposed testing procedure, fifteen main relevant joints consisting of head, neck, left shoulder, right shoulder, shoulder spine, mid spine, base spine, left hip, right hip, left foot, right foot, left knee, right knee, left ankle, and right ankle are tracked over time to create a profile of human body. However, there is no need to index time as Kinect was already set to record 30 fps of all joints in both directions, so the number of points gathered will show how long it takes for someone to walk the 10-foot testing path. Also, among three aforementioned possible axes, only Y (vertical) dimension is considered for further statistical analyses in this study.

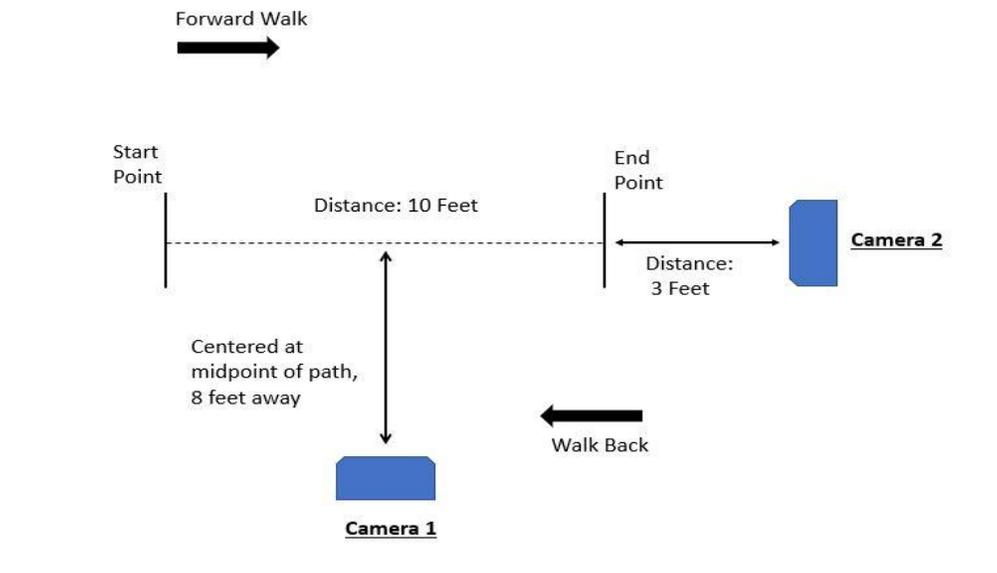

Figure 1: The experimental layout

## 3. Gait (and postural stability) Assessment

The proposed sample entropy method [11] provides a statistic to measure variation in a time series. In this study, we used a revised version of the original SE, called modified SE, details of which can be found in [12]. This section examines the capabilities of Kinect device to produce a profile consisting of modified SE values. In addition, when a subject's gait has deviated from their standard profile, the changes were analyzed. Note that SE value changes can be simply caused either by changing the walking speed or using motion-restricting devices such as braces or walking aid devices. Therefore, we propose to use these devices to study any possible changes in a subject's walking pattern due to using any of the motion-restricting devices. To change a subject's SE value, they were asked to wear motion restricting devices (braces, cane, etc.). We propose to use these devices to study any subsequent changes to a subject's profile.

An experiment was conducted on three male subjects of similar body types and ages. The team equipped test subjects with motion-restricting devices and performed a series of experiments using the proposed testing procedure. All three subjects were asked to walk through the ten-foot path as shown in Figure 1 once wearing a motion-restricting device and another time without the device. For this study, an Ankle Brace and an ACL Brace, were used as the motion-restricting devices. Three trial runs were conducted for each individual with and without each device. The devices were worn on the right side of the patients' body as shown in Figure 2. All tests were completed on the same day.

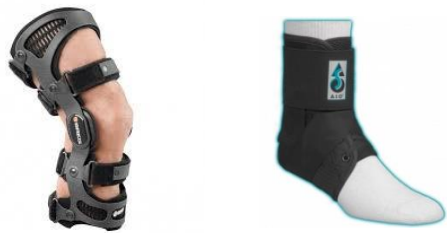

Figure 2: Motion-Restricting Devices (ACL Brace and Ankle Brace)

The position of all relevant joints mentioned in section 2 were gathered for characterizing the postural status of subjects. Malmir and Chang [13] proposed a method using some gait parameters to detect a person's gait change. However, in the current study, raw data is considered to provide better insight and analysis on gait changes. The process of transferring the raw data into a quantitative value for measurement and comparisons goals are described in the next section.

## 4. Quantifying Gait Changes

The data derived from Kinect for each joint makes a profile containing multiple joints. This profile is transformed into numerical values for comparisons. The proposed modified sample entropy measure was applied to quantify variability of the data obtained from both back and forth walking tests into two values. The average of these values was then used to compare different walking conditions. It can help distinguish a subject by their personal gait profile and also identify whether a subject has worn a certain motion-restricting device or not.

Final results for subjects 1, 2, and 3 denoted as *S1*, *S2*, and *S3* are shown in Figure 3 (A), (B), and (C), respectively. The notation of subject wearing knee brace is denoted as *KB* shown as a solid line. The test for ankle brace is denoted as *AB* shown in dotted lines. Finally, the normal walking test is denoted as *NW* shown in dashed lines

$S1 : Subject\ 1$      $KB : Knee\ Brace$
$S2 : Subject\ 2$      $AB : Ankle\ Brace$
$S3 : Subject\ 3$      $NW : Normal\ Walking$

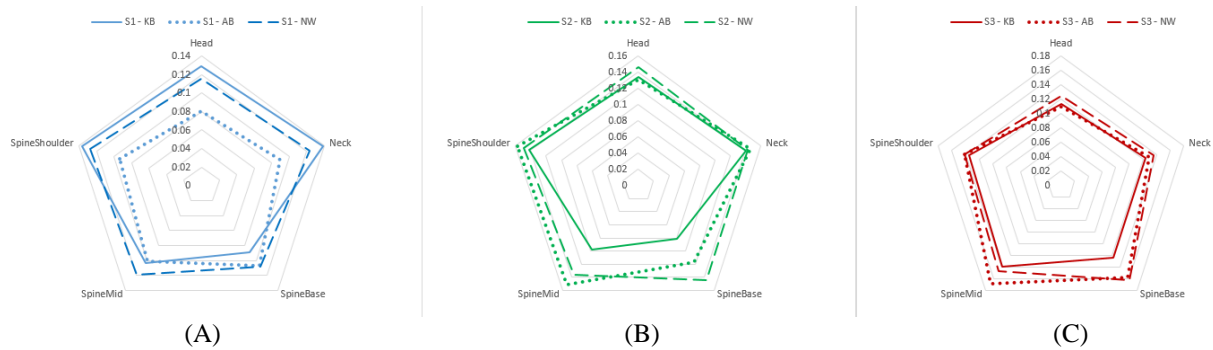

Figure 3: Comparison of the mean SE values of the profiles of five main joints in three different conditions.

Out of 15 joints tracked, a group of upper body joints consisting of head, neck, base spine, mid spine and shoulder spine are primary joints used to quantify changes with the least amount of variance. Figure 3 shows differences between SE values of five main joints for each person in three different conditions. Some SE values are pretty close while some others have significant differences. The variability of SE values in base spine and mid spine in different conditions were larger than the variability of SE values in the other main joints. Therefore, these two joints may be the potential candidates to measure physical therapy progress.

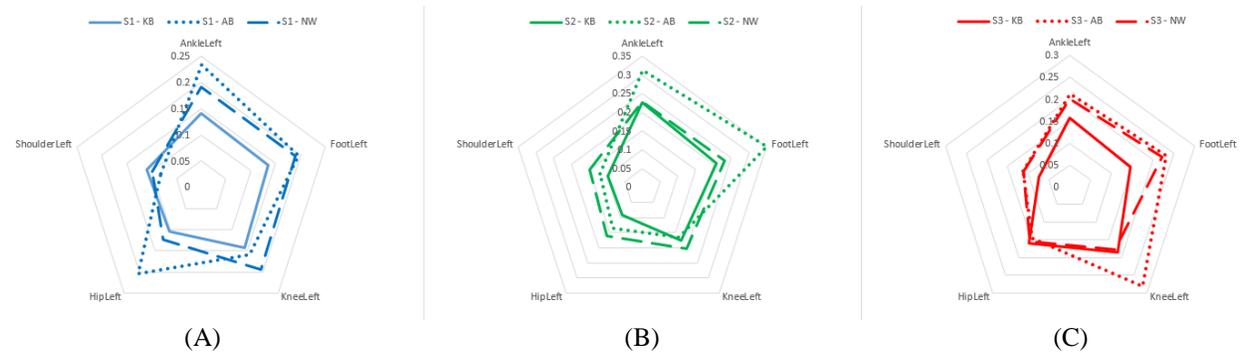

Figure 4: Comparison of the mean SE values of the profiles for five left side joints in three different conditions.

The combinational use of all joints on star glyphs may provide a more concise presentation of gait changes if tracked over time. However, since the braces were worn on the right side, five main joints of the left side, as well as five main joints of the right side of human body were considered for further analysis. All the results discussed so far were obtained from the side camera. The same procedure was done for statistical analyses of the data on vertical dimension derived from the frontal camera as well. The results were similar to those from the side camera in terms of patterns.

Figure 4 illustrates differences between SE values of five left side joints for each person in three different conditions. As seen in the figures, affected joints when each person wears a brace and walks is different in different subjects.

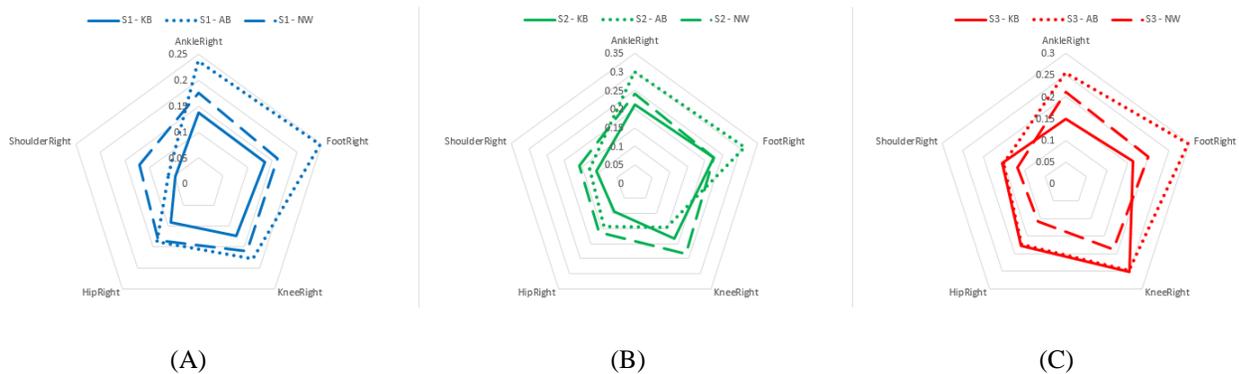

(A)  (B)  (C)

Figure 5: Comparison of the mean SE values of the profiles for five right side joints in three different conditions.

Figure 5 demonstrates the results obtained from the same analysis, but on five right side joints of each individual. Similar to the analysis of the middle body joints, some SE values are pretty close while some others have significant differences. For example, there is almost no differences between the variability of right knee, right hip, and right shoulder of Subject 3 when this subject uses either one of motion-restricting devices. However, he did not have a consistent walking pattern as expected since SE values of his right foot and right ankle were different. Comparing Figures 4 and 5, we see a relatively similar pattern in pentagons related to same five joints in left and right side of each individual's body in all three different conditions. The same scale is considered for both cases to provide fair comparisons.

As seen in Figures 3-5, subjects have different walking patterns. Profiles consisted of SE values from various joints show that motion-restricting devices do alter their gait patterns. We found that more joints should be considered collectively. For example, the joint from right foot does not show much difference for S2 as shown in Figure 5(B) while S2 wore a knee brace. However, the pentagon shapes of KB and NW were much different. The goal in this study was not to find the best joint to compare people's gait but rather to quantify a person's walking patterns so the gait changes can be identified.

Note that there is variability between and within subjects, though it may seem like a possible obstacle it is possible to use this variability to our advantage. The varying SE values within a person are far less than the differences when comparing two separate profiles, this allows us to still properly identify individuals. While looking at the variability within subjects we are focusing on the issues, physical or otherwise, that might be causing the changes. The differences within subjects are the key to allowing us to assist with advancing rehabilitation as well as possibly predicting future injuries.

## 5. Conclusion (or Discussion) and Ongoing Studies

The proposed method relies on a standardized experimentation process to gather relevant data. Kinect raw data was used to compare walking patterns of individuals with almost identical physical characteristics (e.g. gender, height, weight, age). Motion-restricting devices used in everyday life were chosen to validate the usefulness of the proposed system. Initial experimental results confirmed that the proposed method is capable of quantifying differences in walking patterns of different people under different conditions. The proposed sample entropy measure was used to summarize data from each walking path as a profile into one value.

The experimental results demonstrate that wearing the motion-restricting devices alters walking patterns captured by relevant joints in healthy adult males. The exact amount of this change can be quantified simply by using the proposed sample entropy measure [10]. This study validates the hypothesis that the proposed personal profiles for individual subjects can be used to track changes in joints.

All systems have their limitations, ours is no different. The limitations of the proposed method stem mostly from Kinects camera's abilities. Like with all cameras we have restrictions based on lighting and camera angles. We tested in a well-lit room to make sure enough definition is available to the camera. As well as keeping our cameras level at a set height to account for the errors that could span from a changing camera angle. Our system works the best when subjects wear shorts, tennis shoes, and a t-shirt. If a subject wears loose fitting pants the system returns data that is sporadic and therefore unusable. Also, the colors that subjects wear are important, black clothing usually ends up in unusable data. These limitations must be accounted for before we can expand the usage of the proposed system to other environments and situations.

Our future studies may include elderly subjects as the target population. By tracking the walking profiles over time, it opens the doors to identify possible fall related changes. However, we need to consider other motion-restricting devices such as walking cane, and four-legged walking aids which are not included in this study. Another potential future study may be an identification system that can distinguish one subject from another through his/her personal gait profile. This future study requires far more subjects to validate its results.

## References

1. Eltoukhy, M., Kuenze, C., Oh, J., Jacopetti, M., Wooten, S., and Signorile, J. 2017. Microsoft Kinect can distinguish differences in over-ground gait between older persons with and without Parkinson's disease. Medical Engineering and Physics, 44, 1-7.
2. Xu, X., Robertson, M., Chen, K. B., Lin, J. H., and McGorry, R. W. 2017. Using the Microsoft Kinect™ to assess 3-D shoulder kinematics during computer use. Applied Ergonomics.
3. Vernon, S., Paterson, K., Bower, K., McGinley, J., Miller, K., Pua, Y. H., and Clark, R. A. 2015. Quantifying individual components of the timed up and go using Kinect in people living with stroke. Neurorehabilitation and neural repair, 29(1), 48-53.
4. Sosnoff, J. J., Sandroff, B. M., and Motl, R. W. 2012. Quantifying gait abnormalities in persons with multiple sclerosis with minimal disability. Gait and posture, 36(1), 154-156.
5. Hondori, H. M., Khademi, M., 2014, "A Review on Technical and Clinical Impact of Microsoft Kinect on Physical Therapy and Rehabilitation," Journal of Medical Engineering, vol. 2014, Article ID 846514, 16 pages.
6. Malmir, B. (2018). Exploratory studies of human gait changes using depth cameras and sample entropy (Doctoral dissertation).
7. Maki, B. E. 1997. Gait changes in older adults: predictors of falls or indicators of fear?. Journal of the American geriatrics society, 45(3), 313-320.
8. Cham, R., and Redfern, M. S. 2002. Changes in gait when anticipating slippery floors. Gait and posture, 15(2), 159-171.
9. Plotnik, M., Giladi, N., and Hausdorff, J. M. 2007. A new measure for quantifying the bilateral coordination of human gait: effects of aging and Parkinson's disease. Experimental brain research, 181(4), 561-570.
10. Shull, P. B., Jirattigalachote, W., Hunt, M. A., Cutkosky, M. R., and Delp, S. L. 2014. Quantified self and human movement: a review on the clinical impact of wearable sensing and feedback for gait analysis and intervention. Gait and posture, 40(1), 11-19.
11. Ramdani, S., Seigle, B., Lagarde, J., Bouchara, F., and Bernard, P. L. 2009. On the use of sample entropy to analyze human postural sway data. Medical engineering and physics, 31(8), 1023-1031.
12. Koppel, S., and Chang, S. I. 2017. Profile monitoring using Modified Sample Entropy. In IIE Annual Conference. Proceedings (pp. 2015-2020). Institute of Industrial and Systems Engineers (IISE).
13. Malmir, B., & Chang, S. I. (2019). Gait Change Detection Using Parameters Generated from Microsoft Kinect Coordinates. arXiv preprint arXiv:1902.10283.